# The exact solutions and their linear stability analysis for 2-dimensional Ablowitz-Ladik equation[*]

Zhang Jinliang[†], Wang Hongxian

School of Mathematics and Statistics, Henan University of Science and Technology, Luoyang 471023, China

**Abstract:** The Ablowitz-Ladik equation is a very important model in the nonlinear mathematical physics. In this paper, the hyperbolic function solitary wave solutions, the trigonometric function periodic wave solutions and the rational wave solutions with more arbitrary parameters of 2-dimensional Ablowitz-Ladik equation are derived by using the (G'/G )-expansion method, and the effect of the parameters (including the coupling constant and other parameters) on the linear stability of the exact solutions is analysed and numerically simulated.

## 1 Introduction

Many phenomena including nonlinear optics (waveguide arrays), matter waves (Bose-Einstein condensates trapped in optical lattices), plasma physies ( langmuir wave ) and molecular biology (modeling the DNA double strand) are often described by nonlinear partial differential equations (PDEs) [1-3], especially nonlinear Schrodinger-type equations. In the recent years, a number of powerful methods have been proposed, namely the inverse scattering method[3-4], Hirota's bilinear method[6-7], Backlund transformation method[3-5,8], Darboux transformation method[3-4], Hyperbolic function method[9], homogeneous balance method[10-12].

The discrete nonlinear Schrodinger-type equation is a relevant model for a wide range of applications, and has been paid attention by many researchers[13-19] . Dai et al[13,20] used the extended tanh-function approach to derive the exact solutions of discrete complex cubic Ginzburg–Landau equation, discrete complex cubic–quintic Ginzburg–Landau equation with non-local quintic term respectively. Chow et al[15] presented exact solutions of Ablowitz–Ladik model in terms of elliptic functions. Huang & Liu [21] obtained Jacobi elliptic function solutions of the Ablowitz–Ladik system via the Jacobi elliptic function expansion approach. Maruno et al[22] studied a new quintic discrete nonlinear Schrodinger equation by singularity confinement criteria and growth properties. In Ref.[23], the exact localized and periodic solutions of the discrete complex Ginzburg–Landau equation were given, and in Ref.[24], Maruno et al derived a set of exact solutions which includes, as particular cases, bright and dark soliton solutions, constant magnitude solutions with phase shifts, periodic solutions in terms of elliptic Jacobi functions in general forms, and various particular periodic solutions to discrete complex cubic–quintic Ginzburg–Landau equation with a non-local quintic term. Aslan, etc used the $G'/G$ -expansion method to derive the discrete solitons of the discrete nonlinear Schrodinger-type equations [25-26].

*Project supported by the Basic Science and the Front Technology Research Foundation of Henan Province of China (Grant nos. 092300410179, 122102210427), the Doctoral Scientific Research Foundation of Henan University of Science and Technology (Grant no. 09001204) and the Scientific Research Innovation Ability Cultivation Foundation of Henan University of Science and Technology (Grant no.011CX011).

[†] Corresponding author. E-mail: zhangjin6602@163.com

In the practical application, only stable soliton is valuable. Thus the investigation of the soliton stability is very important. Kevrekidis et al use the V-K criterion to analyse the stability of the stationary solutions of higher-dimensional Ablowitz-Ladik model[27]. In the Ref.[28-29], Khare et al consider the saturable descrete nonlinear Schrodinger equation, two-dimensional cubic-quintic discrete nonlinear Schrodinger equation respectively and analyse the stability of the stationary solutions. In the Ref.[30-31], Zhang et al use the (G'/G )-expansion method[32-34] to derive the exact solutions of the discrete complex cubic Ginzburg–Landau equation, the quintic discrete nonlinear Schrodinger equation respectively and study their linear stability.

Due to the important applications in the nonlinear optics, Bose-Einstein condensates, etc, 2–dimensional Ablowitz-Ladik equation is studied in this paper, the exact solutions and their linear stability are researched in the sections 2,3 respectively.

## 2  The exact solutions of (2+1)-dimensional Ablowitz-Ladik equation

In this paper, we consider (2+1)-dimensional Ablowitz-Ladik equation as

$$i\frac{\partial u_{n,m}}{\partial t}+\varepsilon(u_{n+1,m}+u_{n-1,m}+u_{n,m+1}+u_{n,m-1}-4u_{n,m})-\frac{1}{4}|u_{n,m}|^2 \times (u_{n+1,m}+u_{n-1,m}+u_{n,m+1}+u_{n,m-1})=0. \quad (1)$$

According to the (G'/G )-expansion method, the exact solutions of Eq.(1) are derived.

**Case 1**  when $\lambda^2-4\mu>0$, the trigonometric function periodic wave solutions are obtained as following:

$$u_{n,m}(t)=\pm 2\sqrt{\frac{\varepsilon\left[\cos(d_1)\sinh^2(\Delta_1 h_2)+\cos(h_1)\sinh^2(\Delta_1 d_2)\right]}{\left[\cos(d_1)\coth^2(\Delta_1 d_2)\sinh^2(\Delta_1 h_2)+\cos(h_1)\coth^2(\Delta_1 h_2)\sinh^2(\Delta_1 d_2)\right]}}\left(\frac{C_1+C_2\tanh(\Delta_1\xi_{n,m})}{C_1\tanh(\Delta_1\xi_{n,m})+C_2}\right)\times$$

$$\exp\left(i\left(d_1 n+h_1 m+2\varepsilon\left(\frac{(\cos(d_1)\csc h(\Delta_1 d_2)\sinh(\Delta_1 h_2)+\cos(h_1)\csc h(\Delta_1 h_2)\sinh(\Delta_1 d_2))^2}{\cos(d_1)\coth^2(\Delta_1 d_2)\sinh^2(\Delta_1 h_2)+\cos(h_1)\coth^2(\Delta_1 h_2)\sinh^2(\Delta_1 d_2)}-2\right)t+\zeta_1\right)\right)$$

where

$$\xi_{n,m}=d_2 n+h_2 m+\zeta_2-\frac{4\varepsilon\left[\cos(d_1)\sinh^2(\Delta_1 h_2)+\cos(h_1)\sinh^2(\Delta_1 d_2)\right]}{\sqrt{\lambda^2-4\mu}\tanh(\Delta_1 d_2)\tanh(\Delta_1 h_2)}\times$$

$$\frac{\sin(d_1)\tanh(\Delta_1 h_2)+\sin(h_1)\tanh(\Delta_1 d_2)}{\cos(d_1)\coth^2(\Delta_1 d_2)\sinh^2(\Delta_1 h_2)+\cos(h_1)\coth^2(\Delta_1 h_2)\sinh^2(\Delta_1 d_2)}t,$$

$\Delta_1=\frac{\sqrt{\lambda^2-4\mu}}{2}$, $C_1$, $C_2$, $\zeta_1$, $\zeta_2$, $d_2$ and $h_2$ are constants, and $d_1$ 和 $h_1$ satisfy $\cos(d_1)\cos(h_1)>0$.

The exact solutions above are simplified as

$$u_{n,m}^{(1)}(t) = 2\sqrt{\frac{\varepsilon\left[\sinh^2(h)+\sinh^2(d)\right]}{\coth^2(d)\sinh^2(h)+\coth^2(h)\sinh^2(d)}} \frac{C_1+C_2\tanh(dn+hm)}{C_1\tanh(dn+hm)+C_2} \quad (2)$$

$$\times \exp\left(2i\left(\frac{(\csc h(d)\sinh(h)+\csc h(h)\sinh(d))^2}{\coth^2(d)\sinh^2(h)+\coth^2(h)\sinh^2(d)}-2\right)\varepsilon t\right)$$

where $C_1$, $C_2$, $d$ and $h$ are constants, and $n, m \in \mathbb{N}$.

From (2), we can derive two special-type exact solutions of Eq.(1).

**Case 1.1** *Dark soliton*:

$$u_{n,m}^{(11)}(t) = 2\sqrt{\frac{\varepsilon\left[\sinh^2(h)+\sinh^2(d)\right]}{\coth^2(d)\sinh^2(h)+\coth^2(h)\sinh^2(d)}}\tanh(dn+hm) \quad (3)$$

$$\times \exp\left(2i\left(\frac{(\csc h(d)\sinh(h)+\csc h(h)\sinh(d))^2}{\coth^2(d)\sinh^2(h)+\coth^2(h)\sinh^2(d)}-2\right)\varepsilon t\right)$$

where $d$, $h$ are constants.(Figure 1)

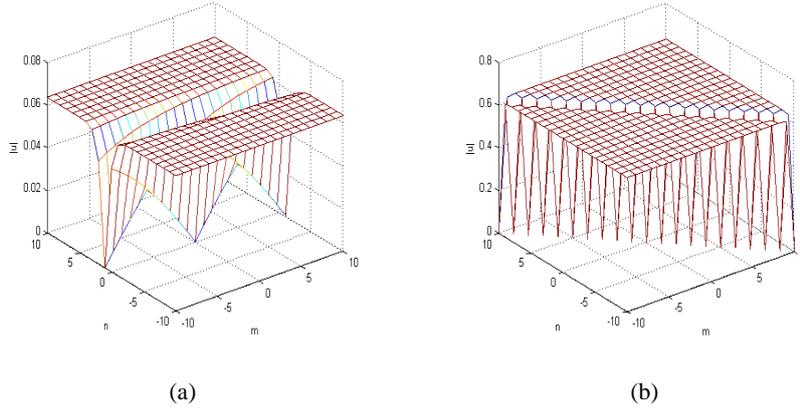

(a)          (b)

Figure 1: Panel (a) Dark soliton (3) when $d=1, h=0.1, \varepsilon=0.1, t=0$

Panel (b) Dark soliton (3) when $d=1, h=1, \varepsilon=0.1, t=0$

**Case 1.2** *Bright soliton* (Figure 2):

$$u_{n,m}^{(12)}(t) = 2\sqrt{\frac{\varepsilon\left[\sinh^2(h)+\sinh^2(d)\right]}{\coth^2(d)\sinh^2(h)+\coth^2(h)\sinh^2(d)}} \frac{2+\tanh(dn+hm)}{2\tanh(dn+hm)+1}$$

$$\times \exp\left(2i\left(\frac{(\csc h(d)\sinh(h)+\csc h(h)\sinh(d))^2}{\coth^2(d)\sinh^2(h)+\coth^2(h)\sinh^2(d)}-2\right)\varepsilon t\right) \quad (4)$$

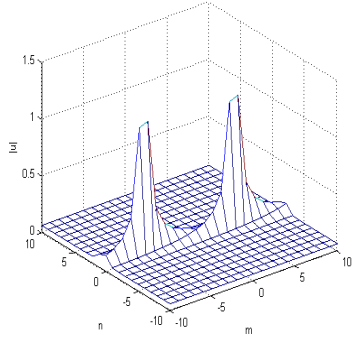

Figure 2: Bright soliton (4) when $d = 1, h = 0.1, \varepsilon = 0.1, t = 0$

**Case 2** when $\lambda^2 - 4\mu < 0$, the trigonometric function periodic wave solutions are obtained as following:

$$u_{n,m}(t) = \pm 2\sqrt{\frac{\varepsilon\left[\cos(d_1)\sin^2(\Delta_2 h_2) + \cos(h_1)\sin^2(\Delta_2 d_2)\right]}{\left[\cos(d_1)\cot^2(\Delta_2 d_2)\sin^2(\Delta_2 h_2) + \cos(h_1)\cot^2(\Delta_2 h_2)\sin^2(\Delta_2 d_2)\right]}} \left(\frac{-C_1 \tan(\Delta_2 \xi_{n,m}) + C_2}{C_1 + C_2 \tan(\Delta_2 \xi_{n,m})}\right) \times$$

$$\exp\left(i\left(d_1 n + h_1 m + 2\varepsilon\left(\frac{(\cos(d_1)\csc(\Delta_2 d_2)\sin(\Delta_2 h_2) + \cos(h_1)\csc(\Delta_2 h_2)\sin(\Delta_2 d_2))^2}{\cos(d_1)\cot^2(\Delta_2 d_2)\sin^2(\Delta_2 h_2) + \cos(h_1)\cot^2(\Delta_2 h_2)\sin^2(\Delta_2 d_2)} - 2\right)t + \zeta_1\right)\right)$$

where

$$\xi_{n,m} = d_2 n + h_2 m + \zeta_2 - \frac{4\varepsilon\left[\cos(d_1)\sin^2(\Delta_2 h_2) + \cos(h_1)\sin^2(\Delta_2 d_2)\right]}{\sqrt{4\mu - \lambda^2}\tan(\Delta_2 d_2)\tan(\Delta_2 h_2)} \times$$

$$\frac{\sin(d_1)\tan(\Delta_2 h_2) + \sin(h_1)\tan(\Delta_2 d_2)}{\cos(d_1)\cot^2(\Delta_2 d_2)\sin^2(\Delta_2 h_2) + \cos(h_1)\cot^2(\Delta_2 h_2)\sin^2(\Delta_2 d_2)} t,$$

$\Delta_2 = \frac{\sqrt{4\mu - \lambda^2}}{2}$, $C_1$, $C_2$, $\zeta_1$, $\zeta_2$, $d_2$ and $h_2$ are constants, and $d_1$ 和 $h_1$ satisfy $\cos(d_1)\cos(h_1) > 0$.

The exact solutions above are simplified as

$$u_{n,m}^{(2)}(t) = -\sqrt{\frac{\varepsilon\left[\sin^2(h) + \cos(h)\sin^2(d)\right]}{\left[\cot^2(d)\sin^2(h) + \cot^2(h)\sin^2(d)\right]}} \left(\frac{-C_1 \tan(dn + hm) + C_2}{C_1 + C_2 \tan(dn + hm)}\right) \times \qquad (5)$$

$$\exp\left(i\left(2\varepsilon\left(\frac{(\csc(d)\sin(h) + \csc(h)\sin(d))^2}{\cot^2(d)\sin^2(h) + \cot^2(h)\sin^2(d)} - 2\right)t\right)\right),$$

where $C_1$, $C_2$, $d$ and $h$ are constants, and $n, m \in \mathbb{N}$.

From (3), we can derive two special-type exact solutions of Eq.(1).

**Case 2.1** *Oscillatory 'tan' solution* (Figure 3):

$$u_{n,m}^{(21)}(t) = 2\sqrt{\frac{\varepsilon\left[\sin^2(h)+\sin^2(d)\right]}{\left[\cot^2(d)\sin^2(h)+\cot^2(h)\sin^2(d)\right]}}\tan(dn+hm)\times$$
$$\exp\left(2i\left(\frac{(\csc(d)\sin(h)+\csc(h)\sin(d))^2}{\cot^2(d)\sin^2(h)+\cot^2(h)\sin^2(d)}-2\right)\varepsilon t\right) \quad (6)$$

where $d$, $h$ are constants.

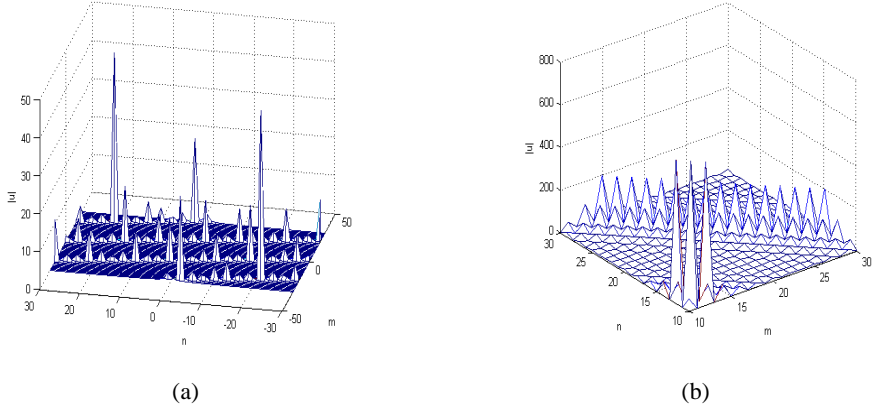

(a)        (b)

Figure 3: Panel (a) Oscillatory 'tan' solution (6) when $d=3.2, h=3.3, \varepsilon=0.1, t=0$

Panel (b) Oscillatory 'tan' solution (6) when $d=1.5, h=1.5, \varepsilon=0.1, t=0$

**Case 2.2** *The trigonometric function periodic wave solutions* ( Figure 4)：

$$u_{n,m}^{(22)}(t) = -2\sqrt{\frac{\varepsilon\left[\sin^2(h)+\sin^2(d)\right]}{\left[\cot^2(d)\sin^2(h)+\cot^2(h)\sin^2(d)\right]}}\frac{-\tan(dn+hm)+2}{1+2\tan(dn+hm)}\times$$
$$\exp\left(2i\left(\frac{(\csc(d)\sin(h)+\csc(h)\sin(d))^2}{\cot^2(d)\sin^2(h)+\cot^2(h)\sin^2(d)}-2\right)\varepsilon t\right) \quad (7)$$

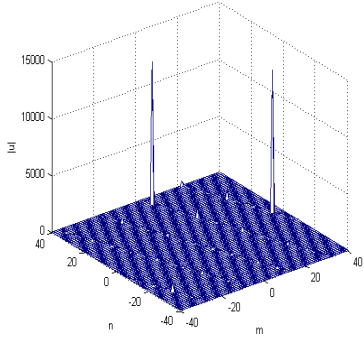

Figure 4: *The trigonometric function periodic wave solutions* (7)

when $d=3.2, h=3.3, \varepsilon=0.1, t=0$

**Case 3** when $\lambda^2 - 4\mu = 0$

$$u_{n,m}(t) = \pm 2d_2 h_2 \sqrt{\frac{\varepsilon\left(d_2^2\cos(h_1)+h_2^2\cos(d_1)\right)}{\left(d_2^4\cos(h_1)+h_2^4\cos(d_1)\right)}}\left(\frac{C_1}{C_1\xi_{n,m}+C_2}\right)\times$$

$$\exp\left(i\left(d_1 n + h_1 m + 2\varepsilon\left(\frac{\left(d_2^2 \cos(h_1) + h_2^2 \cos(d_1)\right)^2}{d_2^4 \cos(h_1) + h_2^4 \cos(d_1)} - 2\right)t + \zeta_1\right)\right)$$

where $\xi_{n,m} = d_2 n + h_2 m - \dfrac{2\varepsilon d_2 h_2 \left(d_2^2 \cos(h_1) + h_2^2 \cos(d_1)\right)\left(d_2 \sin(h_1) + h_2 \sin(d_1)\right)}{d_2^4 \cos(h_1) + h_2^4 \cos(d_1)} t + \zeta_2$,

$C_1, C_2, \zeta_1, \zeta_2, d_2$ and $h_2$ are constants, and $d_1$ 和 $h_1$ satisfy $\cos(d_1)\cos(h_1) > 0$.

The exact solution above is simplified as (Figure 5)

$$u_{n,m}^{(3)}(t) = 2dh\sqrt{\frac{\varepsilon(d^2+h^2)}{(d^4+h^4)}}\left(\frac{1}{dn+hm+0.5}\right)\exp\left(i\left(2\varepsilon\left(\frac{(d^2+h^2)^2}{d^4+h^4}-2\right)t\right)\right) \quad (8)$$

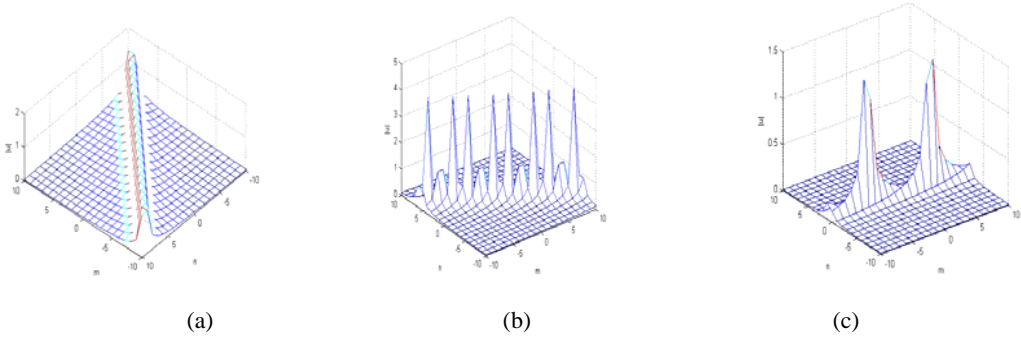

(a)          (b)          (c)

Figure 5 : Panel (a) *Rational wave solution* (8) when $d=1, h=1, \varepsilon=0.1, t=0$

Panel (b) *Rational wave solution* (8) when $d=1, h=0.6, \varepsilon=0.1, t=0$

Panel (c) *Rational wave solution* (8) when $d=1, h=0.11, \varepsilon=0.1, t=0$

**Note:** The other type exact solutions of Eq.(1) can be obtained if the parameters are setted appropriately.

## 3   The linear stability analysis of the exact solutions for (2+1)-dimensional Ablowitz-Ladik equation

In order to analyse the linear stability of exact solutions $u_{n,m}^{(j)}(j=1,2,3)$, we suppose

$$u_{n,m} = u_{n,m}^{(j)} e^{-i\omega t} + \delta u_{n,m}(t) e^{-i\omega t} \quad (9)$$

where $\delta u_{n,m}(t)$ is a small perturbation. Substituting (9) into Eq.(1) and retaining only terms linear in the perturbation yields

$$i\delta \dot{u}_{n,m} + \left(\varepsilon - \frac{1}{4}\left|u_{n,m}^{(j)}\right|^2\right)\left(\delta u_{n+1,m} + \delta u_{n-1,m} + \delta u_{n,m+1} + \delta u_{n,m-1}\right) + (\omega - 4\varepsilon)\delta u_{n,m}$$
$$-\frac{1}{4}u_{n,m}^{(j)}\left(\delta u^*_{n,m} + \delta u_{n,m}\right)\left(u_{n+1,m}^{(j)} + u_{n-1,m}^{(j)} + u_{n,m+1}^{(j)} + u_{n,m-1}^{(j)}\right) = 0 \quad (10)$$

Setting $\delta u_{n,m} = \delta u^r_{n,m} + i\delta u^i_{n,m}$, Eq.(10) can be changed into

$$\delta \dot{u}^r + (\omega - 4\varepsilon)\delta u^i_{n,m} + \left(\varepsilon - \frac{1}{4}\left|u^{(j)}_{n,m}\right|^2\right)\left(\delta u^i_{n+1,m} + \delta u^i_{n-1,m} + \delta u^i_{n,m+1} + \delta u^i_{n,m-1}\right) = 0 \tag{11}$$

$$-\delta \dot{u}^i_{n,m} + \left[\omega - 4\varepsilon - \frac{1}{2}u^{(j)}_{n,m}\left(u^{(j)}_{n+1,m} + u^{(j)}_{n-1,m} + u^{(j)}_{n,m+1} + u^{(j)}_{n,m-1}\right)\right]\delta u^r_{n,m}$$
$$+ \left(\varepsilon - \frac{1}{4}\left|u^{(j)}_{n,m}\right|^2\right)\left(\delta u^r_{n+1,m} + \delta u^r_{n-1,m} + \delta u^r_{n,m+1} + \delta u^r_{n,m-1}\right) = 0 \tag{12}$$

Supposing $J = n + (m-1)N$, and introducing the two real vectors

$$\delta U^r = \{\delta u^r_J\}, \quad \delta U^i = \{\delta u^i_J\},$$

where $N$ and $M$ are periodical parameters of solitons, and the two real matrices

$$\mathbf{A} = \{A_{J,J'}\}, \quad \mathbf{B} = \{B_{J,J'}\};$$

where

$$A_{J,J'} = (\omega - 4\varepsilon)\delta_{J,J'} + \left(\varepsilon - \frac{1}{4}\left|u^{(j)}_{n,m}\right|^2\right)\left(\delta_{J,J'-1} + \delta_{J,J'+1} + \delta_{J,J'-N} + \delta_{J,J'+N}\right)$$
$$= (\omega - 4\varepsilon)\delta_{J,J'} + \left(\varepsilon - \frac{1}{4}\left|u^{(j)}_J\right|^2\right)\left(\delta_{J,J'-1} + \delta_{J,J'+1} + \delta_{J,J'-N} + \delta_{J,J'+N}\right) \tag{13}$$

$$B_{J,J'} = \left[\omega - 4\varepsilon - \frac{1}{2}u^{(j)}_{n,m}\left(u^{(j)}_{n+1,m} + u^{(j)}_{n-1,m} + u^{(j)}_{n,m+1} + u^{(j)}_{n,m-1}\right)\right]\delta_{J,J'}$$
$$+ \left(\varepsilon - \frac{1}{4}\left|u^{(j)}_{n,m}\right|^2\right)\left(\delta_{J,J'-1} + \delta_{J,J'+1} + \delta_{J,J'-N} + \delta_{J,J'+N}\right) \tag{14}$$
$$+ \left(\varepsilon - \frac{1}{4}\left|u^{(j)}_J\right|^2\right)\left(\delta_{J,J'-1} + \delta_{J,J'+1} + \delta_{J,J'-N} + \delta_{J,J'+N}\right)$$

where $J' \pm 1$ in the Kronecker $\delta$ means: $J' \pm 1 \bmod N$, and $J' \pm N$ $J' \pm 1$ in the Kronecker $\delta$ means: $J' \pm N \bmod NM$. Then Eqs. (11)–(12) can be written compactly as

$$\delta \dot{U}^r + \mathbf{A}\delta U^i = 0, \quad -\delta \dot{U}^i + \mathbf{B}\delta U^r = 0 \tag{15}$$

From (15), we have

$$\delta \ddot{U}^r + \mathbf{AB}\delta U^r = 0, \quad \delta \ddot{U}^i + \mathbf{BA}\delta U^i = 0 \tag{16}$$

The eigenvalue spectrum of the matrices **AB** and **BA** determines the stability of the exact solutions. If all the eigenvalues are positive, the solutions are stable, and if eigenvalues contain a negative (or complex) eigenvalue, the solution is unstable.

### 3.1 The linear stability analysis of the hyperbolic function solitary wave solutions

### 3.1.1 The effect of the parameters $d$ and $h$ on the linear stability of $u^{(11)}_{n,m}$

From the Fig. 6, it is easy to see that the stability region of $u^{(11)}_{n,m}$ is along the coordinate axis of $d$ and $h$, and is symmetrical to line $d = h$ when $M = N$. When $N = 3, M = 3$, $u_{n,m}^{(11)}$ is linear stable in the region of $d > 1, h < 0.5$ and $h > 1, d < 0.5$. With the increment of $N$ and $M$, the stability region becomes small.

With the increment of $N$ and $M$, the stability region ($d < 1, h < 1$) is discretized. This is similar to the cubic-quintic discrete nonlinear Schrodinger equation[22].

Similar to the analysis to $u_{n,m}^{(11)}$, the effect of the parameters $d$ and $h$ on the linear stability region of $u_{n,m}^{(12)}$ is shown in Fig. 7 when $\varepsilon=0.1, N = 3, M = 3$ (the black regions are the stability region)

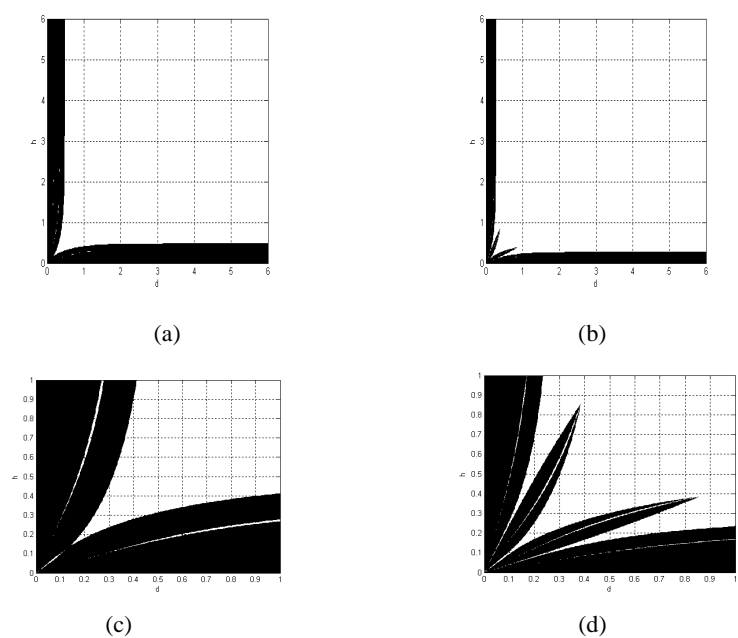

(a) (b)

(c) (d)

Figure 6: The effect of the parameters $d$ and $h$ on the linear stability of $u_{n,m}^{(11)}$

Panel (a) $N = 3, M = 3, \varepsilon=0.1$; Panel (b) $N = 5, M = 5, \varepsilon=0.1$ (the black regions are the stable region)

Panel (c) partial enlargement drawing of Panel (a); Panel (d) partial enlargement drawing of Panel (b)

(the black regions are the stable region)

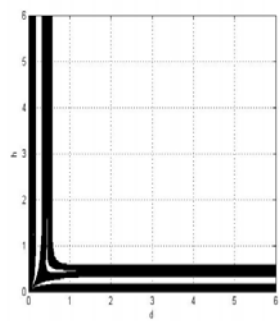

Figure 7. The effect of the parameters $d$ and $h$ on the linear stability of $u_{n,m}^{(12)}$ when $\varepsilon=0.1, N = 3, M = 3$

### 3.1.2 The effect of the parameters $C_1$ and $C_2$ on the linear stability of $u_{n,m}^{(1)}$

For the sake of simplicity, we suppose that parameters $C_1$ and $C_2$ are nonnegative and set $(C_1, C_2) = (\sin\theta, \cos\theta)$ and $\theta \in \left[0, \dfrac{\pi}{2}\right]$. The distribution graphs of the minimal eigenvalues of **AB** of the exact solutions with the changing of the parameters $C_1$ and $C_2$ are shown in Fig.8.

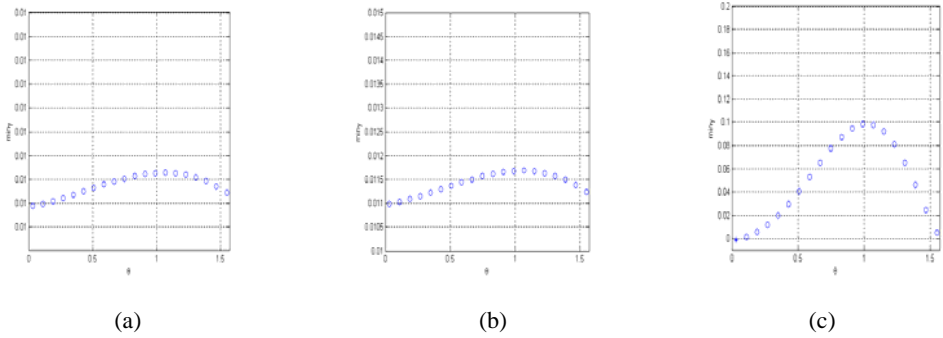

(a)  (b)  (c)

Figure 8 The distribution graphs of the minimal eigenvalues of $AB$ when $N=3$, $M=3$, $\varepsilon=0.1$, and

Panel (a) $d=1, h=0.01$; Panel (b) $d=1, h=0.1$; Panel (c) $d=1, h=1$

($O$ -Positive eigenvalue; $*$-Negative eigenvalue; $\nabla$ -complex eigenvalue)

### 3.1.3 The effect of the coupling constant $\varepsilon$ on the linear stability of $u_{n,m}^{(11)}$

The coupling constant $\varepsilon$ is very important to the Ablowitz-Ladik system, and when $\varepsilon \to 0$, Ablowitz-Ladik system is changed to discrete limiting form and the coupling constant $\varepsilon$ influences the stability of the solutions [20].

The distribution graph of the minimal eigenvalues of **AB** of the exact solutions with the changing of the coupling constant $\varepsilon$ is shown in Fig.9. From Fig.9, it is easy to obtain that the coupling constant $\varepsilon$ has not influence on the stability of the solution $u_{n,m}^{(11)}$ in essence except for strength of the stability of the solution $u_{n,m}^{(11)}$.

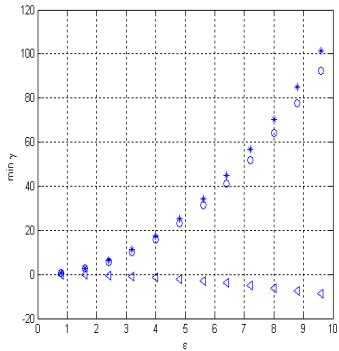

Figure 9. The distribution graphs of the minimal eigenvalues of $AB$ when $N=3$, $M=3$,

Case(a) $d=1, h=0.01(o)$; Case (b) $d=1, h=0.1(*)$; Case (c) $d=1, h=1(\triangleleft)$

## 3.2 The linear stability analysis of the trigonometric function periodic wave solutions

### 3.2.1 The effect of the parameters $d$ and $h$ on the linear stability of $u_{n,m}^{(21)}$

From Fig.10, it is easy to obtain that the stability regions of $u_{n,m}^{(21)}$ is massive and discretized, and with the increment of $N$ and $M$, the stability region becomes small, and the instable points appear in the massive stability region. The large massive stability regions appear in the neighbour of the points $(d,h)=(3n,3n)$. By using the numerical simulation, we have that $u_{n,m}^{(21)}$ is instable in the point $(d,h)=(3.2,3.3)$ if $N>10, M>10$.

The stability region almost exclusively appears in the neighbour of the points $(d,h)=(3,3)$ if $N \neq M$ (Fig. 11) and the stability region remarkably reduces. This is similar to the two-dimensional cubic-quintic discrete nonlinear Schrodinger equation[22].

The effect of the parameters $d$ and $h$ on the stability region of $u_{n,m}^{(22)}$ when $N=3, M=3$ is shown in Fig. 12.

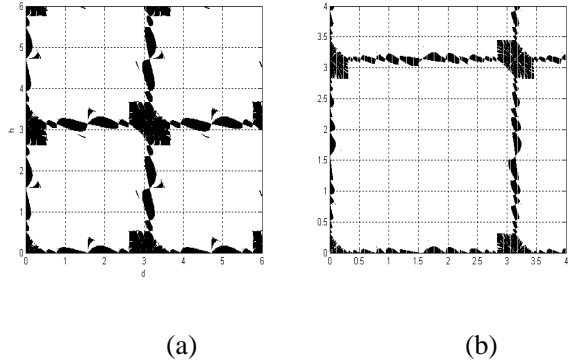

(a)　　　　　　　(b)

Figure 10: The effect of the parameters $d$ and $h$ on the linear stability of $u_{n,m}^{(21)}$

Panel (a) $N=3, M=3$; Panel (b) $N=5, M=5$ (the black regions are the stable region)

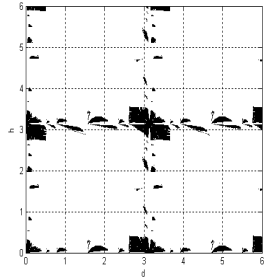

Figure 11. The effect of the parameters $d$ and $h$ on the linear stability of $u_{n,m}^{(21)}$ when $N=3, M=4$

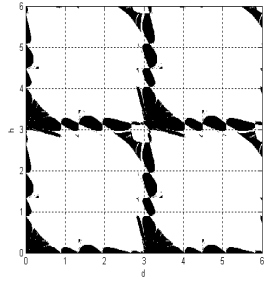

Figure 12. The effect of the parameters $d$ and $h$ on the linear stability of $u_{n,m}^{(22)}$ when $N=3, M=3$

### 3.2.2 The effect of the parameters $C_1$ and $C_2$ on the linear stability of $u_{n,m}^{(2)}$

Similar to Section 3.1.2, The distribution graphs of the minimal eigenvalues of **AB** with the changing of the parameters $C_1$ and $C_2$ are shown in Fig.13.

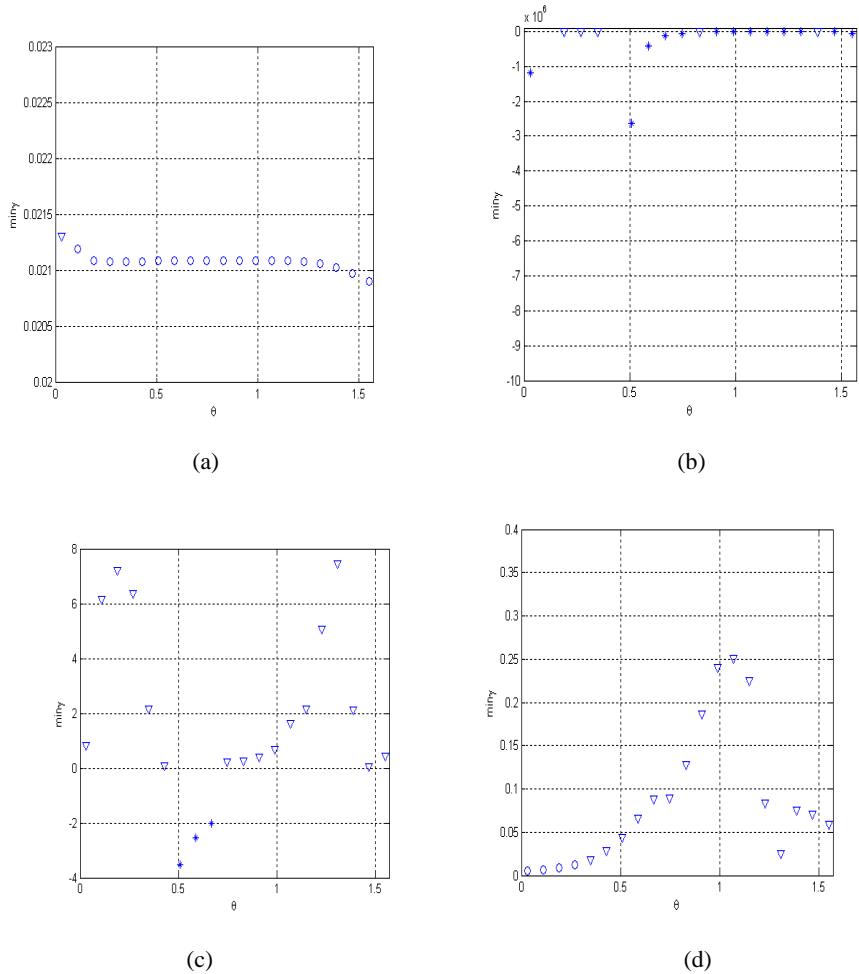

Figure 13. The distribution graphs of the minimal eigenvalues of $AB$ when $N=3$, $M=3$, $\varepsilon=0.1$, and Panel (a) $d=3.2, h=3.3$; Panel (b) $d=1.5, h=1.5$; Panel (c) $d=1, h=1$; Panel (d) $d=2.8, h=2.8$
( $O$ -Positive eigenvalue; $*$ -Negative eigenvalue; $\nabla$ -complex eigenvalue)

### 3.2.3 The effect of the coupling constant $\varepsilon$ on the linear stability of $u_{n,m}^{(21)}$

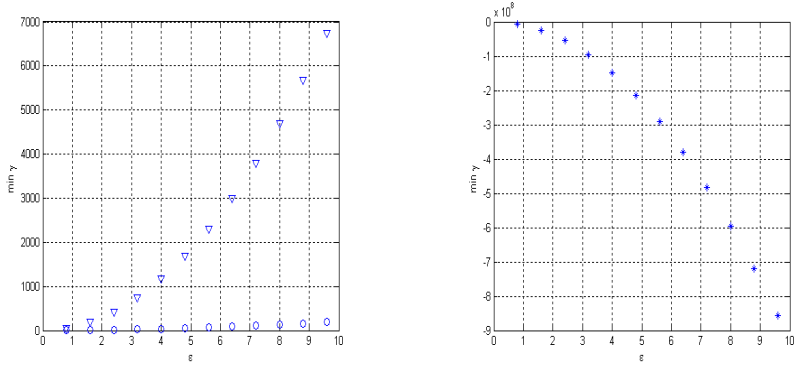

Figure 14. The distribution graphs of the minimal eigenvalues of *AB* when $N = 3$, $M = 3$,
Case(a) $d = 3.2$, $h = 3.3(o)$; Case(b) $d = 1, h = 0.1(\nabla)$; Case(c) $d = 1.5$, $h = 1,5$ (*)

The distribution graphs of the minimal eigenvalues of **AB** of the exact solutions with the changing of the coupling constant $\varepsilon$ are shown in Fig.14. It is similar to Section 3.1.3.

### 3.3 The linear stability analysis of the rational wave solution $u_{n,m}^{(3)}$

We use the method above to analyse the linear stability of $u_{n,m}^{(3)}$. From Fig. 15, the stability region of $u_{n,m}^{(3)}$ is more larger than the other solutions, and the stability regions are discretized with the increment of $N$ and $M$; when $M \to \infty, N \to \infty$, $u_{n,m}^{(3)}$ is linear stable in the neighbour of line $d = h$.

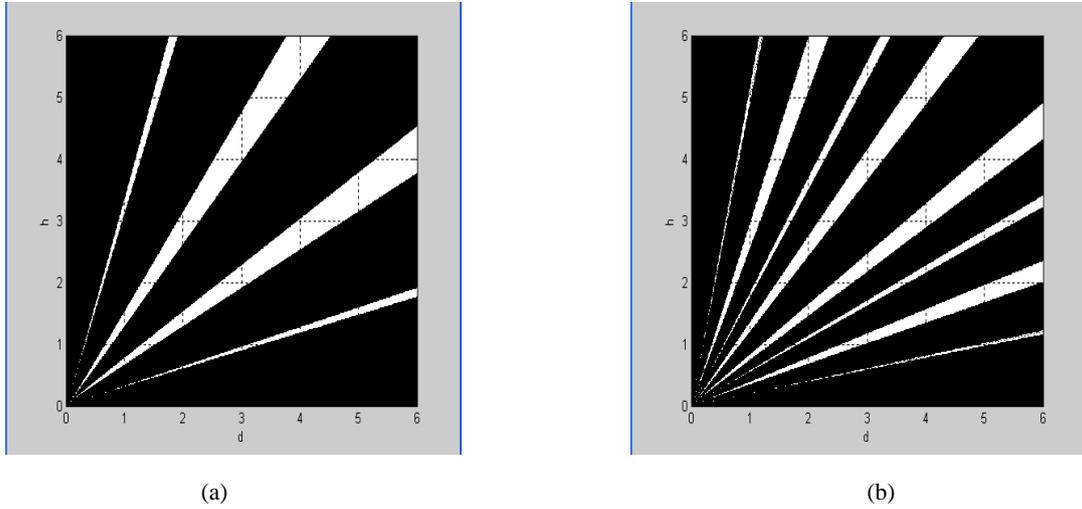

(a)          (b)

Figure 15. The effect of the parameters $d$ and $h$ on the linear stability of $u_{n,m}^{(3)}$
Panel (a) $N = 3, M = 3$; Panel (b) $N = 5, M = 5$

The distribution graph of the minimal eigenvalues of **AB** of the exact solutions with the changing of the coupling constant $\varepsilon$ is shown in Fig.16. It is similar to Section 3.1.3.

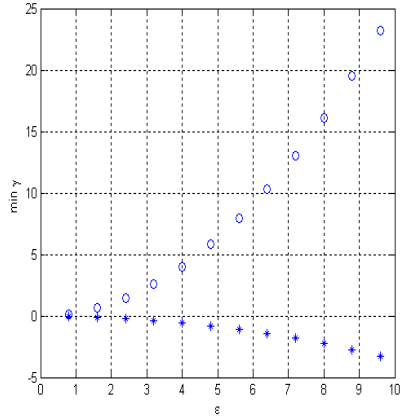

Figure 16. The distribution graphs of the minimal eigenvalues of *AB* when $N=3$, $M=3$,

Case (a) $d=3$, $h=3$ (o), Case(b) $d=2$, $h=3$ (*)

## 4  Conclusions and Discussions

In this paper, the hyperbolic function solitary wave solutions, the trigonometric function periodic wave solutions and the rational wave solutions with more arbitrary parameters of 2-dimensional Ablowitz-Ladik equation are derived using the (G'/G )-expansion method, and the effect of the parameters on the linear stability of the exact solutions is discussed. From the analysis and numerical simulation, we obtain:

(1) The coupling constant $\varepsilon$ has not influence on the stability of the solutions includinig the bright solitons, dark solitons and rational wave solutions except for strength of the stability.
(2) With the increment of $N$ and $M$, the stability region of the solutions includinig the bright solitons, dark solitons and rational wave solutions is discretized if the parameters $d$ and $h$ are considered.

**References：**
[1] Kevrekidis P G 2009 *The discrete nonlinear Schrodinger equation: Mathematical analysis, numerical computations and physical perspectives* (Berlin: Springer-Verlag)p3
    Kevrekidis P G 2011 *IMA Journal of Applied Mathematics* **76** 389
[2] Kuznetsov E A, Rubenchik A M, Zakharov V E 1986 *Physics Reports* **142** 103
[3] Ablowitz M J and Clarkson P A 1991 *Solitons, Nonlinear Evolution Equations and Inverse Scattering*(New York:Cambridge University Press) p200
[4] Gu C H, et al 1990 *Soliton Theory and its Application* (Hangzhou: Zhejiang Publishing House of Science and Technology) p160
[5] Miura M R 1978 *Backlund Transformation* (Berlin: Springer-Verlag) p185
[6] Hirota R 1971 *Phys. Rev. Lett.* **27** 1192
[7] Hu X B, Tam H W 2000 *Physics Letters A* **276** 65
[8] Sun M N, Deng S F, Chen D Y 2005 *Chaos, Solitons and Fractals* **23** 1169
[9] Parkes E J, Duffy B R 1996 *Computer Physics Commutions* **98** 288
[10] Wang M L 1995 *Phys. Lett. A* **199** 169
[11] Wang M L, Zhou Y B, Li Z B 1996 *Physics Letters A* **216** 67
[12] Zhang J L, Wang Y M, Wang M L, Fang Z D 2003 *Chinese Physics* **12** 245

[13] Dai C Q, Cen X, Wu S S 2008 *Computers & Mathematics with Applications* **56** 55
[14] Soto-Crespo J M, Akhmediev N, Ankiewicz A 2003 *Phys. Lett. A* **314** 126
[15] Chow K W, Conte R, Xu N 2006 *Phys. Lett. A* **349** 422
[16] Malomed B, Weinstein M I 1996 *Phys. Lett. A* **220** 91
[17] Malomed B A, Crasovan L C, Mihalache D 2002 *Physica D* **161** 187
[18] Pelinovsky D E, Kevrekidis P G, Frantzeskakis D J 2005 *Physica D* **212** 1
[19] Chong C, Pelinovsky D E 2011 *Discrete and Continuous Dynamical Systems, Series S* **4** 1019
[20] Dai C Q andZhang J F 2006 *Optics Communications* **263** 309
[21] Huang W H and Liu Y L 2009 *Chaos, Solitons and Fractals* **40** 786
[22] Maruno K, Ohta Y and Joshi N 2003 *Physics Letters A* **311** 214
[23] Maruno K, Ankiewicz A and Akhmedievc N 2003 *Optics Communications* **221** 199
[24] Maruno K, Ankiewicz A and Akhmedievc N 2005 *Physics Letters A* **347** 231
[25] Aslan I 2011 *Physics Letters A* **375** 4214
[26] Aslan I 2009 *Applied Mathematics and Computation* **215** 3140
[27] Kevrekidis P G, Herring G J, Lafortune S, Hoq Q E 2012 *Physics Letters A* **376** 982
[28] Khare A, Rasmussen K Ø, Samuelsen M R, Saxena A 2005 *Journal of Physics A: Mathematical and General* **38** 807
[29] Khare A, Rasmussen K Ø, Samuelsen M R, Saxena A 2011 *Physica Scripta* **84** 065001.
[30] Zhang J L, Liu Z G 2011 *Communications of the theoretical physics* **56** 1111
[31] Zhang J L, Liu Z G, Li S W and Wang M L 2012 *Physica Scripta* **86** 015401.
[32] Wang M L, Li X Z, Zhang J L2008 *Phys. Lett. A* **372** 417
[33] Wang M L, Zhang J L, Li X Z 2008 *Applied Mathematics and Computation* **206** 321
[34] Guo S M, Zhou Y B 2010 *Applied Mathematics and Computation* **215** 3214